\documentclass[12pt,preprint]{aastex}
\usepackage{colordvi}
\usepackage{color}
\begin{document}

\title{Ultra-narrow Negative Flare Front Observed in Helium-10830~\AA\ using the 1.6 m New Solar Telescope}

\author{Yan Xu\altaffilmark{1,2},
        Wenda Cao\altaffilmark{2},
        Mingde Ding\altaffilmark{3},
        Lucia Kleint\altaffilmark{4},
        Jiangtao Su\altaffilmark{5,6},
        Chang Liu\altaffilmark{1,2},
        Haisheng Ji\altaffilmark{7},
        Jongchul Chae\altaffilmark{8},
        Ju Jing\altaffilmark{1,2},
        Kyuhyoun Cho\altaffilmark{8},
        Kyungsuk Cho\altaffilmark{9},
        Dale Gary\altaffilmark{2},
        and
        Haimin Wang\altaffilmark{1,2}}

\affil{1.\ Space Weather Research Lab, Center for
Solar-Terrestrial Research, \\ New Jersey Institute of Technology \\
323 Martin Luther King Blvd, Newark, NJ 07102-1982}

\affil{2.\ Big Bear Solar Observatory, \\ New Jersey Institute of Technology \\
323 Martin Luther King Blvd, Newark, NJ 07102-1982}

\affil{3.\ School of Astronomy and Space Science, Nanjing University, Nanjing 210093, China}

\affil{4.\ Fachhochschule Nordwestschweiz (FHNW), \\ Institute of 4D technologies \\
Bahnhofstr. 6, CH-5210 Windisch, Switzerland}

\affil{5.\ Key Laboratory of Solar Activity, \\
National Astronomical Observatories, Chinese Academy of Sciences, \\
Beijing 100012, China}

\affil{6.\ State Key Laboratory of Space Weather, \\
Chinese Academy of Sciences, \\
Beijing 100190, China}

\affil{7.\ Purple Mountain Observatory, \\
2 Beijing Xi Lu, Nanjing, 210008, China}

\affil{8.\ Astronomy Program, Department of Physics and Astronomy, \\
Seoul National University, Seoul 151-747, Republic of Korea}

\affil{9.\ Korea Astronomy and Space Science Institute, \\
Daedeokdae-ro 776, Yuseong-gu, Daejeon 305-348, Republic of Korea}

\date{\today}

\clearpage

\begin{abstract}
  Solar flares are sudden flashes of brightness on the Sun and are often associated with coronal mass ejections and solar energetic particles which have adverse effects in the near Earth environment. By definition, flares are usually referred to \textbf{bright} features resulting from excess emission. Using the newly commissioned 1.6~m New Solar Telescope at Big Bear Solar Observatory, here we show a striking ``negative'' flare with a narrow, but unambiguous ``\textbf{dark}'' moving front observed in \ion{He}{1} 10830~\AA, which is as narrow as 340 km and is associated with distinct spectral characteristics in H$\alpha$ and \ion{Mg}{2} lines. Theoretically, such negative contrast in \ion{He}{1} 10830~\AA\ can be produced under special circumstances, by nonthermal-electron collisions, or photoionization followed by recombination. Our discovery, made possible due to unprecedented spatial resolution, confirms the presence of the required plasma conditions and provides unique information in understanding the energy release and radiative transfer in astronomical objects.
\end{abstract}


\section{Introduction}

Solar flares are the result of sudden energy release, often exceeding $10^{32}$ ergs, from magnetic reconnection \citep{Hudson2011}. The main manifestation of flares is extensive emission over much of the electromagnetic spectrum. Intriguingly, under some special circumstances, negative flare contrasts (decrease of intensity) are occasionally reported in stellar observations in visible continuum wavelengths \citep{Flesch1974}. In those cases, a 20\% negative contrast in the optical continuum is typically detected prior to the initial brightening \citep{Hawley1995}. The widely accepted postulate for generating negative flares is the H$^{-}$ absorption model \citep{Grinin1973,Grinin1983}, in which a  bombarding beam of electrons penetrates to the lower and cooler atmosphere and causes enhanced collisional ionization of hydrogen, and leads to an increase of electron density and eventually an increase of H$^-$ opacity. As a result, the formation layer of the visible continuum, though still in the photosphere, is shifted slightly upward and becomes cooler.

Analogously, intensity drops associated with flares can occur on the Sun, but involving a variety of features and formation mechanisms. The continuum dimming caused by the enhanced H$^-$ opacity was rarely observed and the diminution effects are weaker on the Sun than that of the stellar events \citep{Henoux1990}. So far only two events have been reported, with negative contrasts of 5\% at 5500~\AA\ \citep{Henoux1990} and 1\% $\sim$ 2\% at 8500~\AA\ \citep{Ding2003b}. However, due to the measurement uncertainty and low resolution, both events are regarded as non-compelling detections \citep{Ding2003b, Henoux1990}.

In microwave observations, negative contrasts of solar radiation during flares have also been reported \citep{Sawyer1977, Maksimov1991}. The term ``negative flare'' was used but the dimming in radio flux was caused by absorption of intervening filament material, or a similar mechanism, rather than a change in the local emission process itself.

Besides the visible continuum and microwave observations, negative flares can also be found in helium-line observations. Solar observations by \citet{Zirin1980} attracted attention to the \ion{He}{1} D3 line at 5876~\AA, by which the helium element was first discovered in 1868. \citet{Liu2013} confirmed the occurrence of enhanced D3 darkening and discussed fine structures of the dark-bright pattern of flare ribbons. Owing to electrons orbiting the nucleus, a \ion{He}{1} atom comprises two different kinds of energy states: the singlet (para-helium) and the triplet (ortho-helium) states. Since the ground state is a singlet one ($1s^2$ $^1S$), radiative transitions to the higher triplet states ($2p$~$^{3}P$ and $3d$~$^{3}D$) responsible for the D3 line are forbidden. These states are then populated through collisional transitions or photoionization from the ground state followed by recombinations, which require energy from thermal (or nonthermal) electrons and EUV irradiation, respectively. Whether the D3 line appears in emission or absorption depends on whether the line source function in the line forming region is larger or smaller than the nearby continuum intensity. The line source function is in turn determined by the ratio of the populations at the two energy states \citep{Rutten2003}. In the case that these energy states are populated mainly through thermal collisional transitions, the plasma temperature is the key factor in determining whether the D3 line appears in absorption or emission \citep{Centeno2008, Mauas2005, Zirin1988}.

Similar to the D3 line, another helium line, at 10830~\AA, is formed by the transition between $2s$~$^{3}S$ and $2p$~$^{3}P$ of the helium triplet. A negative flare was reported by \citet{Harvey1984}, in \ion{He}{1} 10830~\AA\ using spectrographic data. However, due to the lack of spatial information, the possibility of enhanced absorption of plage or dark filament material moving into the spectral slit can not be ruled out. Two mechanisms,  namely photoionization-recombination and collisional ionization-recombination, that can populate the helium triplet and thus enable the transitions between $2s$~$^{3}S$ and $2p$~$^{3}P$ states, were studied by \citet{Ding2005}. As expected, the EUV radiation turns the \ion{He}{1} 10830 line to emission via photoionization-recombination, which has been confirmed by \citep{Zeng2014}. Notably, the collisional ionization-recombination can enhance the absorption at the beginning of the flare and generate stronger emission at flare maximum than the photoionization-recombination effect \citep{Ding2005}. As such, under collisional bombardment, the intensity in a flaring region could first decrease and then increase afterward. In this study, we present the very first, unambiguous evidence of negative contrast observed during solar flares and discuss the morphological and spectral properties of the negative flares.

\section{Observation}

The key observations in \ion{He}{1} 10830~\AA\ presented in this study, were obtained using the newly commissioned 1.6-m New Solar Telescope \citep[NST;][]{Goode2012} at Big Bear Solar Observatory. Equipped with a high-order adaptive optics system using 308 sub-apertures \citep{Cao2010}, NST provides unprecedented high spatio-temporal resolution. For instance, the image scale at 10830~\AA\ achieves 0\arcsec.085 (61.6 km) per pixel. The near infrared (10830~\AA) imaging system is based on a liquid-nitrogen-cooled Hg\,Cd\,Te focal plane CMOS array and a Lyot-filter \citep{Cao2010}, which was tuned to the blue wing of the \ion{He}{1} line at 10830.05~\AA\ with a passband of 0.5~\AA. The typical field-of-view (FOV) is about 85\arcsec\ and the effective cadence is 15 s after performing the speckle reconstruction \citep{Woger2007}. Supportive H$\alpha$ imaging spectroscopic data were obtained simultaneously by the Fast Imaging Solar Spectrograph \citep[FISS;][]{Chae2013}, which has a smaller FOV of 40\arcsec\ by 60\arcsec\ with a scanning cadence as fast as 20 s and a spectral resolving power ($\Delta\lambda$/$\lambda$) of $1.4 \times 10^5$. Complementary full-disk intensity maps are obtained from two major instruments onboard the Solar Dynamics Observatory \citep[SDO;][]{SDO}, the Helioseismic and Magnetic Imager \citep[HMI;][]{HMI} provided the visible continuum maps and the Atmospheric Imaging Assembly \citep[AIA;][]{AIA} provided UV/EUV maps. The spatial resolution of SDO/HMI and SDO/AIA is lower than that of the NST, but they provide context information in locating the NST target on the solar disk and comparing flare emissions in different wavelengths.

In the past several decades, taking advantage of rapid developments of focal plane instrumentation, substantial flare observations have been carried out by 2-D imaging to provide spatially resolved information. However, there was no high-resolution imaging spectroscopy for solar flares in the visible or UV wavelengths from space \citep{Fletcher2011}. The newly launched NASA mission, IRIS \citep{IRIS}, fills this gap by providing imaging spectroscopy in both near ultraviolet (NUV) and far ultraviolet (FUV) bands. The NUV window of IRIS has a main passband at 2800~\AA, and the FUV window covers the spectra near 1400~\AA. The IRIS raster size is chosen depending on the program. The spectrograph may scan a FOV of 130\arcsec\ by 175\arcsec\ using a slit with dimension of 0.\arcsec33 by 175\arcsec\ for slow observations \citep{IRIS} and may speed up with 1-16 step rasters for flares or faster programs. Pre-flare spectra served as reference and for the study of pre-flare conditions. The slit position is visible in the Slit-Jaw images as a dark vertical line.

\section{Data Analysis and Result}

It is notable that previous \ion{He}{1} D3 observations of solar flares are rare and the data were recorded on film \citep{Liu2013, Zirin1980}. To our knowledge, no negative flare has ever before been clearly recorded using an imaging system in 10830~\AA, probably due to the low sensitivity of detectors and coarse observing resolutions (spatial and temporal). Using the unprecedented 100-km resolution in \ion{He}{1} 10830~\AA, we report the first imaging observation of negative flares. As listed in Table~\ref{nf_list}, two events were observed, on 2013 August 17 (M1.4) and 2014 August 1 (M1.5), and presented in this study.

\begin{table}[pht]
\caption{Negative flares in \ion{He}{1} 10830~\AA, observed by NST. \label{nf_list}}

\begin{tabular}{lccccr}
\\
\tableline\tableline
Date         & GOES SXR  & NST 10830~\AA & FISS/NST H$\alpha$ & IRIS & RHESSI \\
             & magnitude &               &                    &      &        \\

2013 Aug. 17 & M1.4      & Yes           & Yes                & No   & No     \\

2014 Aug. 01 & M1.5      & Yes           & No                 & Yes  & No     \\

\tableline
\end{tabular}
\end{table}

\subsection{Morphological Analysis of the event on 2013 August 17}

\begin{figure}
\centering
\includegraphics[scale=1]{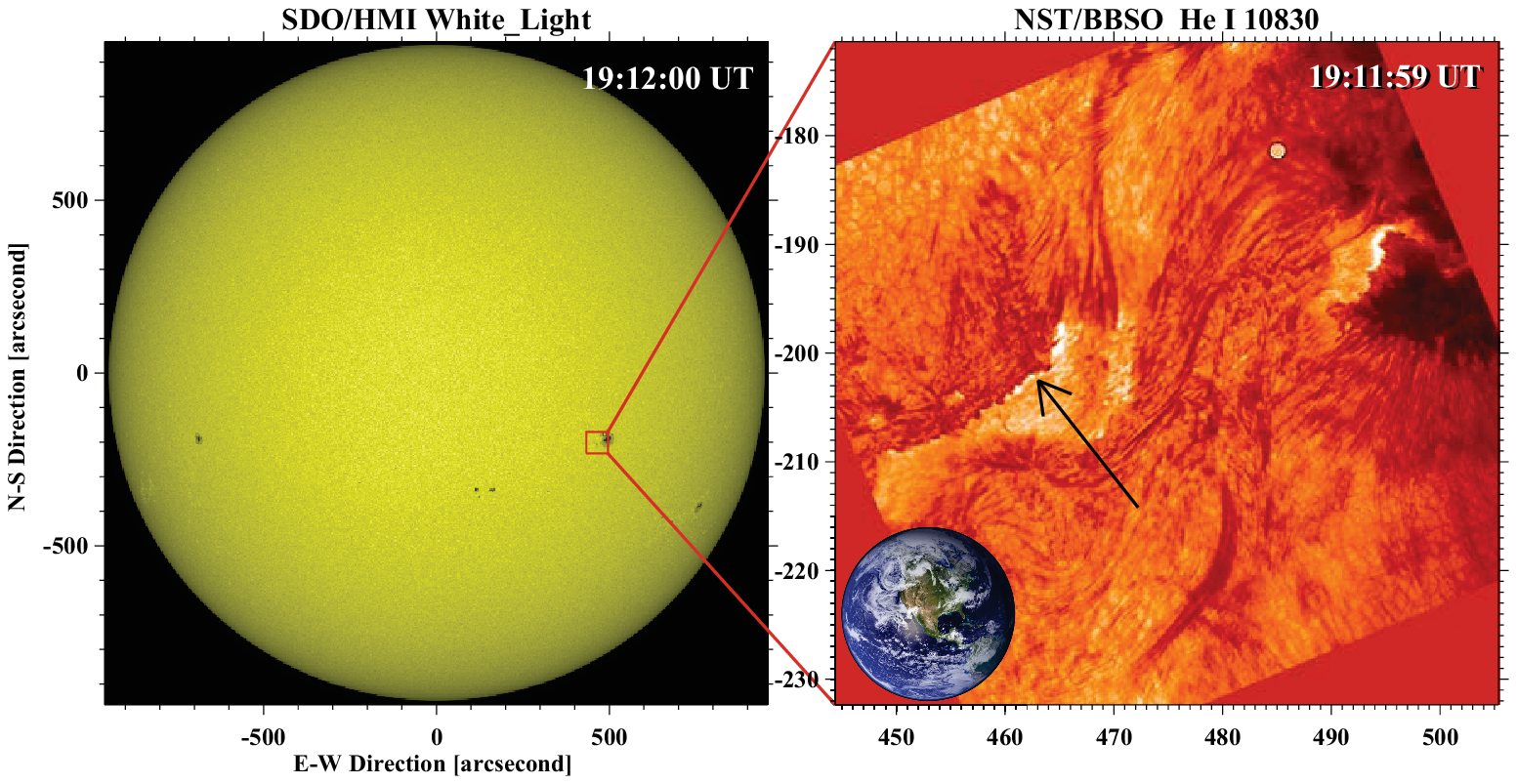}
\caption{Left: The full-disk visible continuum map taken at 19:12:00 UT by SDO/HMI. It serves as the reference to register the NST's FOV with heliographic coordinates (Stonyhurst, see \citep{Thompson2006}), in which the central meridian of solar disk seen from the Earth is set to zero and the unit is in arcsecond ($\sim$ 725 km). The active region NOAA 11818 locates to the southwest from the disk center. The red box outlines the area of interest (AOI) where the negative flare appeared. Right: NST 10830~\AA\ observation taken at 19:11:59 UT registered with the SDO/HMI map in heliographic coordinates. The black arrow points to the flare ribbon of interest with a dark frontier. An image of the Earth is displayed in the corner for comparison of size. The spot on the upper right corner masks a group of hot pixels.
\label{fulldisk}}
\end{figure}

On 2013 August 17, we observed a solar flare with GOES soft-X-ray-class M1.4 at about 18:43 UT in active region NOAA 11818. A filament eruption, or possibly eruption of a magnetic flux rope, was associated with this flare and led to a halo coronal mass ejection (CME). However, the filament disappeared early in the flare period and therefore has no effect on the \ion{He}{1} 10830 dark flare ribbon found much later at the flare footpoints. Figure~\ref{fulldisk} presents the HMI and NST images taken around 19:12 UT showing a negative flare ribbon in the \ion{He}{1} 10830~\AA\ indicated by the black arrow. In the left panel, a full-disk HMI continuum map is displayed, on which a small area is highlighted by a red box corresponding to the NST's FOV. The right panel shows the NST \ion{He}{1} 10830~\AA\ map. This flare has a typical two-ribbon morphology. One ribbon is located inside the sunspot penumbra-umbra region with a stronger magnetic field and the other is located in the surrounding granular area with a weaker magnetic field. We focus on the ribbon outside the sunspot as it is the one associated most clearly with negative contrasts. It is notable that all the NST images are speckle reconstructed to achieve diffraction-limited resolution. The speckle reconstruction could in principle add artificial features to a region with a strong gradient in brightness. In order to confirm that the negative flare signature is real, we verified that the dark ribbon is also clearly identified in the raw data prior to speckle reconstruction. Therefore, we are confident that the observed flare ribbon with a negative contrast is not an artificial feature introduced by the speckle reconstruction.

\begin{figure}
\centering
\includegraphics[scale=1]{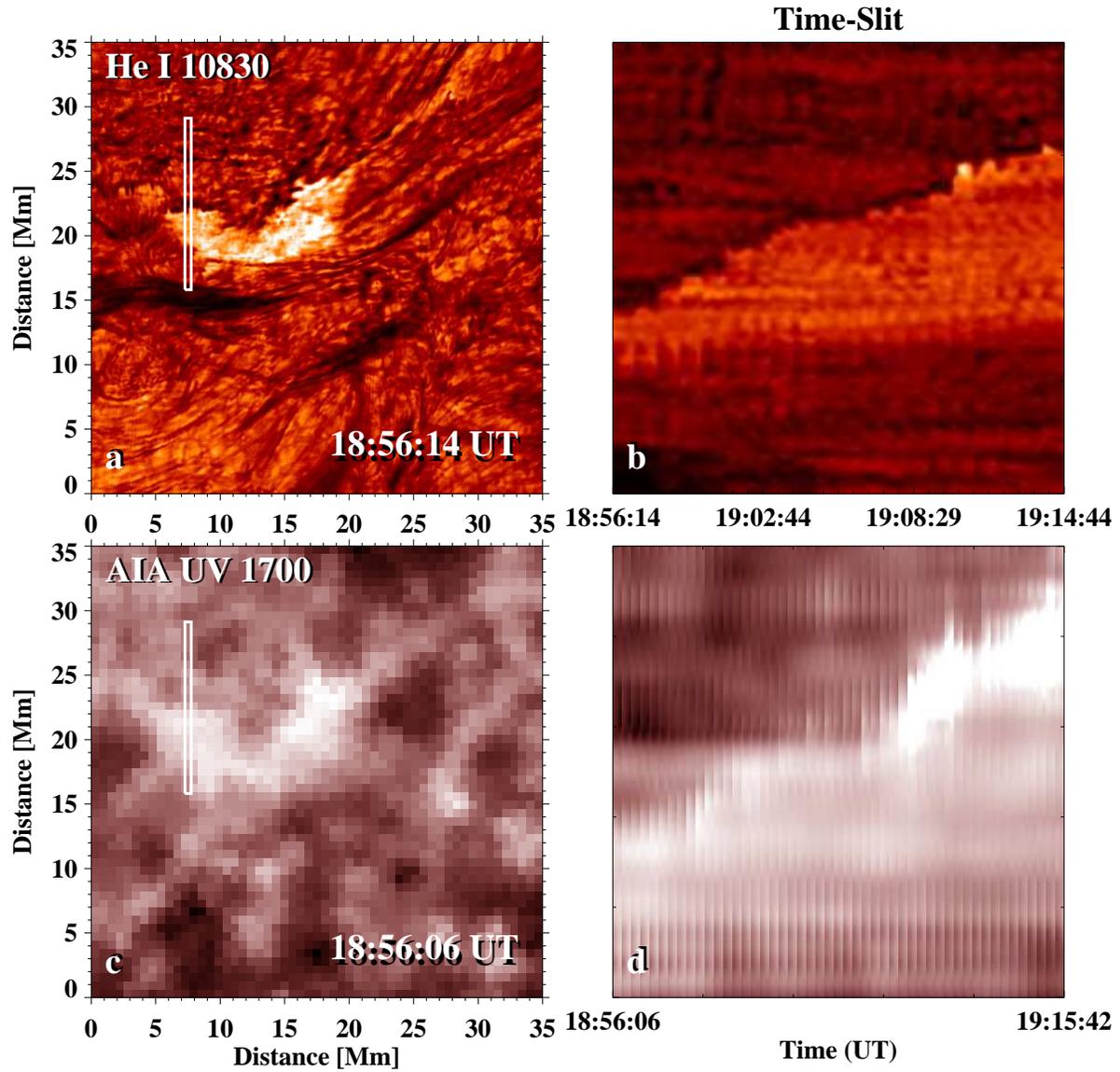}
\caption{a: \ion{He}{1} 10830~\AA\ image taken at 18:56:14 UT. This flare ribbon is basically moving Northward (upward on the image). A vertical slit is selected cutting across the flare ribbon. b: The time-space diagram of the slit. The time period is 1110s, from 18:56:14 UT to 19:14:44 UT, during which the dark edge traveled about 4100 km indicating an average velocity of 3.7 km s$^{-1}$. c: SDO/AIA UV 1700~\AA\ image with a pixel resolution of 0.\arcsec6, taken at 18:56:06 UT. d: UV 1700~\AA\ time-space diagram. It is clear that there is no dark ribbon seen in the UV 1700~\AA\ images.
\label{spacetime}}
\end{figure}

\begin{figure}
\centering
\includegraphics{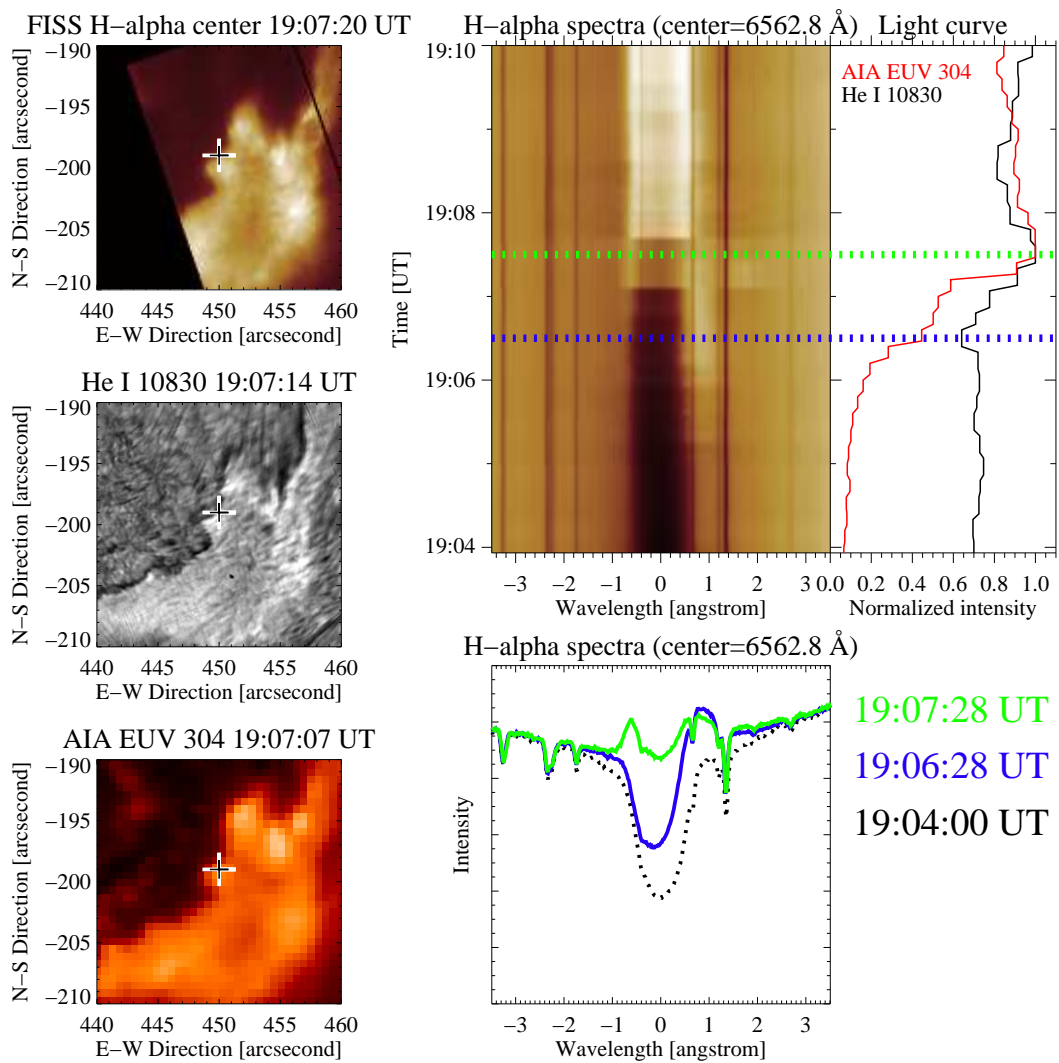}
\caption{Left column (top to bottom): FISS raster image at the H$\alpha$ line center, \ion{He}{1} 10830~\AA\ image and SDO/AIA 304~\AA\ image taken almost simultaneously. The cross located on the dark flare ribbon indicates the position where the H$\alpha$ spectra are obtained. Upper right panel: H$\alpha$ spectra and the light curves of \ion{He}{1} 10830~\AA\ and SDO/AIA 304~\AA\ of the cross center. The H$\alpha$ line profiles at 19:06:28 UT and 19:07:28 UT (see the blue and green lines) are plotted in the same colors in the lower right panel. Lower right panel: H$\alpha$ line profile at three different times, representing the pre-flare (19:04:00 UT), electron precipitation (19:06:28 UT) and evaporation (19:07:28 UT) stages at the location of the cross.
\label{ha}}
\end{figure}

One example of the \ion{He}{1} 10830~\AA\ image taken at 18:56:14 UT is shown in Figure~\ref{spacetime}a. The bright flare source is clearly seen. Owing to NST's unprecedented high resolution, a dark edge is clearly visible on the upper boundary of this flare ribbon. Note that the ribbon is moving upward in the FOV and thus the dark edge is actually the moving front that represents the footpoints of newly reconnected flare loops. In other words, the ribbon with a negative contrast corresponds to the latest electron beams penetrating from the solar corona, where the magnetic reconnection occurred. The bright part of the ribbon behind the dark frontier is the signal of the flare heating immediately following the enhanced line absorption. Such a leading front due to electron penetration has been reported in H$\alpha$ observations as narrow bright features with red-shifted spectra, for instance by \citet{Svestka1980} using the multi-slit spectrograph developed by Lockheed Solar Observatory \citep{Martin1974}. During our observing run, using the imaging spectroscopy by FISS, we were able to identify the red-shift in H$\alpha$ spectra at the same location where the enhanced 10830 absorption occurs. For illustration, as shown in Figure~\ref{ha}, a cross is put on a representative location of the ribbon front, which shows emission in H$\alpha$ and EUV 304~\AA, but absorption in \ion{He}{1} 10830~\AA.  On the upper right panel, the H$\alpha$ spectra on the cross and the light curves of \ion{He}{1} 10830~\AA\ and EUV 304~\AA\ are plotted. At an early time of 19:04:00 UT, the flare emission had not yet propagated to the cross and a typical broad H$\alpha$ absorption line is seen (black dotted curve in the lower right panel). Later on, at 19:06:28 UT, the ribbon-front had arrived at the cross location, indicating the start of precipitation characterized by a strong red-wing enhancement in the  H$\alpha$ line profile (blue curve in the lower right panel). After 19:07:28 UT, the emission dominates in all wavelengths and we see an emission profile in H$\alpha$ with a typical central reversal, resulting from self-absorption \citep{Ding1997a}. This strong red wing emission in H$\alpha$, appearing in the early phase of the flare, indicates the existence of large scale mass motions caused by impulsive heating, consistent with the most recent radiative hydrodynamic simulation of an electron beam heated atmosphere \citep{dacosta2015}. Therefore, a close relationship is built up between the negative flare ribbon and the precipitation of energetic particles.

The above description of the event agrees with the scenario predicted by \citet{Ding2005}, in which the electron beams bombarding on cooler plasma over-populate the $2s$~$^{3}S$ state via collisional ionization-recombination, resulting in the absorption in \ion{He}{1} 10830. Later on, when this layer of the solar atmosphere is efficiently heated to a higher temperature, the $2p$~$^{3}P$ state is also sufficiently populated and emission is produced due to both photoionization-recombination and collisional ionization-recombination, by enhanced EUV irradiation and nonthermal electrons, respectively. The transition from absorption to emission is likely a result of the line source function varying from a lower value (in a cooler atmosphere) to a higher one (in a hotter atmosphere). We select a slit cutting through the dark ribbon (see Figure~\ref{spacetime}a) and plot its time-space diagram in  Figure~\ref{spacetime}b. As such, the motion of the dark ribbon and the expansion of the following bright flare ribbon are clearly identified. By measuring the distance traveled within the time period from 18:56:14 UT to 19:14:44 UT, the average speed of the ribbon motion is calculated to be 3.7 km s$^{-1}$, which is significantly slower than some strong flares \citep[e.g.,][]{Xu2006} but within a typical range for moderate flares. In the bottom panels of Figure~\ref{spacetime}, we see an emission area in UV 1700~\AA, which is almost identical to \ion{He}{1} 10830~\AA\ but obviously without the dark fringe.

\begin{figure}
\centering
\includegraphics[scale=1.1]{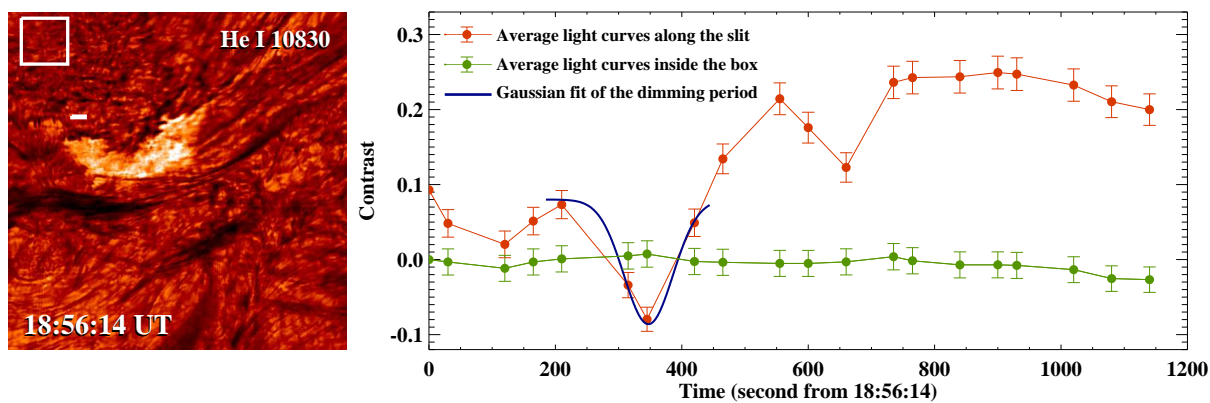}
\caption{Left: \ion{He}{1} 10830~\AA\ image taken at 18:56:14 UT, which corresponds to the starting point of the light curve plotted in the right panel. A horizontal slit is selected in the ribbon's travel path. The temporal evolution of the averaged intensity of the slit reflects the fluctuation caused by the flare ribbon (with both negative and positive contrasts) moving across the slit. The larger box to the upper left corner is an area of quiet Sun, away from the flare. The intensity variation in this region serves as the reference compared with the flare light curve in the small slit. Right: Average light curves inside the box (green) and along the slit (red). The red curve with 1$\sigma$ is the light curve of the flare, from which the sudden dip represents the perturbation of leading dark edge. Please note that the second dip is not a dimming but a relative weak emission. The blue Gaussian curve provides an estimate of the length of time that the dark front spends passing across the slit. The estimate of the maximum decrease in intensity is about $-13.7\%$, comparing to the pre-flare condition.
\label{lc}}
\end{figure}

\begin{figure}
\centering
\includegraphics[scale=1.1]{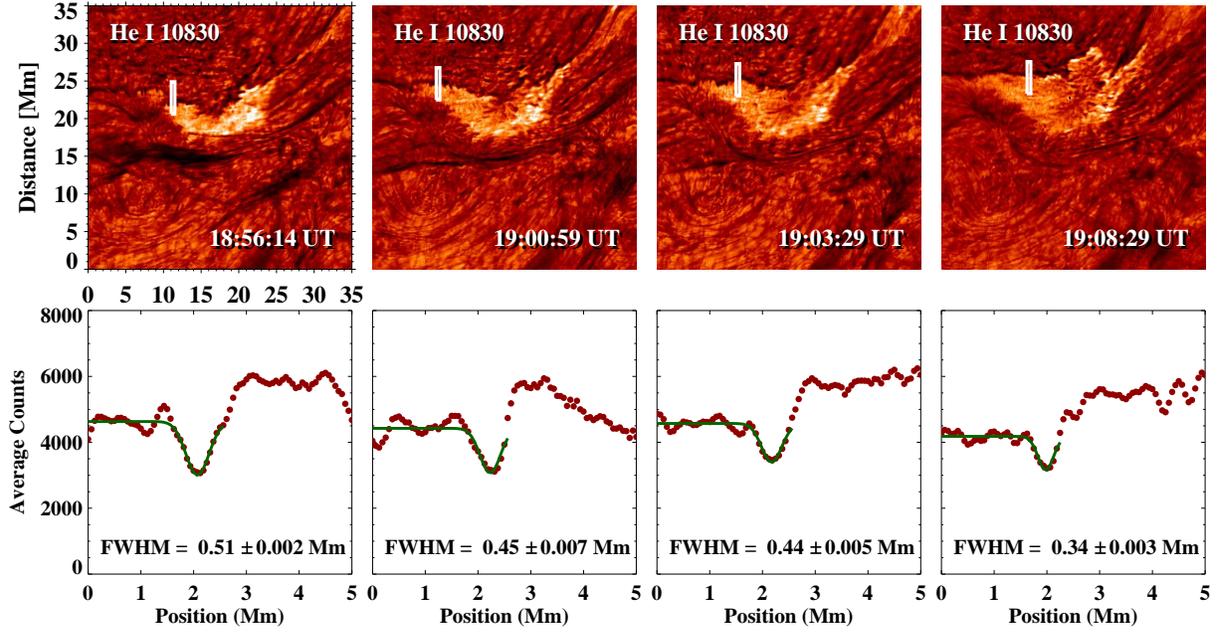}
\caption{Top: Four \ion{He}{1} 10830~\AA\ frames taken at different times are displayed with a slit overplotted on each image. Bottom: The intensity profiles (red) across the slits in top panels are fitted using a Gaussian function (green) plus a constant representing the pre-flare background. In principal, the FWHM is a good proxy of the width (or size) of the dark flare ribbon. The fitting results provide a range of the width from 340 to 510 km. Please note that the fitting results are slightly affected by the selection of background.
\label{width}}
\end{figure}

The temporal variation of intensity is illustrated in Figure~\ref{lc}. On the left panel, a small slit on the path of the dark ribbon is chosen, and a square region is selected as a reference area. The light curves of the averaged intensity in the slit and the large box are plotted in the right panel in red and green colors, respectively. As expected, when the ribbon sweeps through the slit, we see an obvious drop of the intensity compared to the initial stage. Afterwards, the slit region is dominated by the subsequent flare heating and thus a steady increase of the 10830~\AA\ intensity is observed. The error bars represent 1$\sigma$ deviation of the measurements of the contrasts in both the reference area and the slit. Therefore, we are confident that the intensity variation, especially the decrease, is due to neither the variation of the background (green curve) nor the seeing fluctuation. The brightness, relative to the reference area, of the slit area drops from about 5.7\% to $-8.0\%$. This can be translated to a negative contrast of $-13.7\%$ relative to the pre-flare condition. The slit light curve also provides the time scale of the dark flare ribbon. A Gaussian fit of the dimming period shows a full-width-at-half-maximum (FWHM) of about 91 s. Another important parameter is the characteristic size (width of the ribbon), which is crucial for the estimation of electron flux used in the numerical simulation of flare absorption and emission. We measure the size following the method of \citet{Xu2012a}. As shown in Figure~\ref{width}, four slits are selected at different positions and different times across the dark flare ribbon. In the bottom panels, the intensity spatial profile of each slit is fitted using a Gaussian function plus a constant at the pre-flare level. The FWHM measurements give a width of the dark ribbon in a range from 340 km to 510 km. A wider range of the width is possible by selecting more slit positions. However, the four-point measurements are within the best seeing conditions and therefore are more representatively meaningful. Note that the dark flare ribbon is fully resolved during the entire period of observation, and thus the lower limit of the width is at least greater than 120 km (two pixels).

\subsection{IRIS imaging spectroscopy of the event on 2014 August 1}

In the second event on 2014 August 1, the Interface Region Imaging Spectrograph (IRIS) \citep{IRIS} observed the same region as NST. The high resolution UV spectra provide supportive information with respect to the negative flare ribbon observed by NST. While not as striking as the earlier event, the main features of the negative contrast front are the same in the two events.

\begin{figure}
\centering
\includegraphics[scale=1.1]{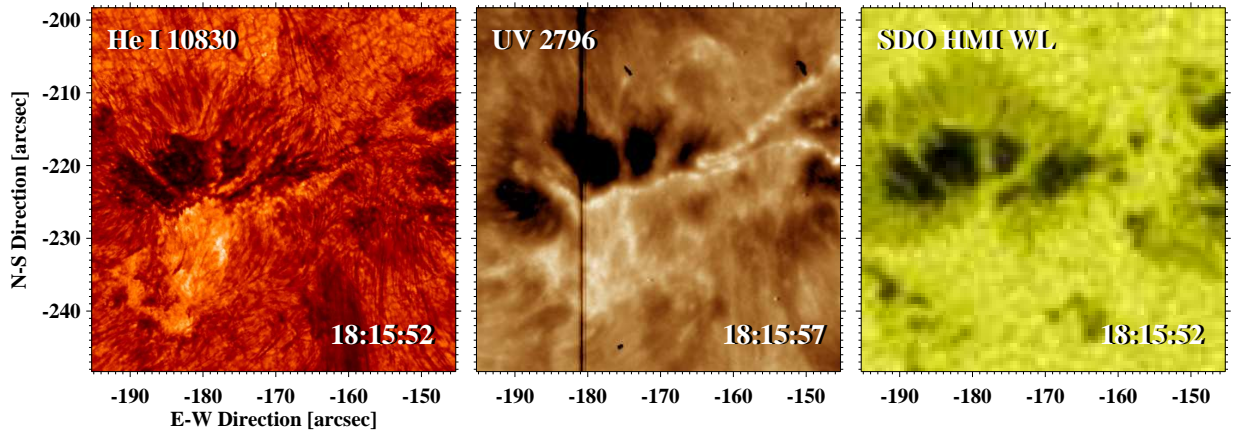}
\caption{Multi-wavelength observations of the negative flare observed on 2014 August 1. Left: NST 10830~\AA\ image with dark ribbon front. Middle: IRIS slit-jaw image at UV 2796~\AA. The vertical dark line shows the slit position at 18:15:17 UT. Right: SDO intensity map used for alignment purpose.
\label{2ndsji}}
\end{figure}

\begin{figure}
\centering
\includegraphics[scale=1.1]{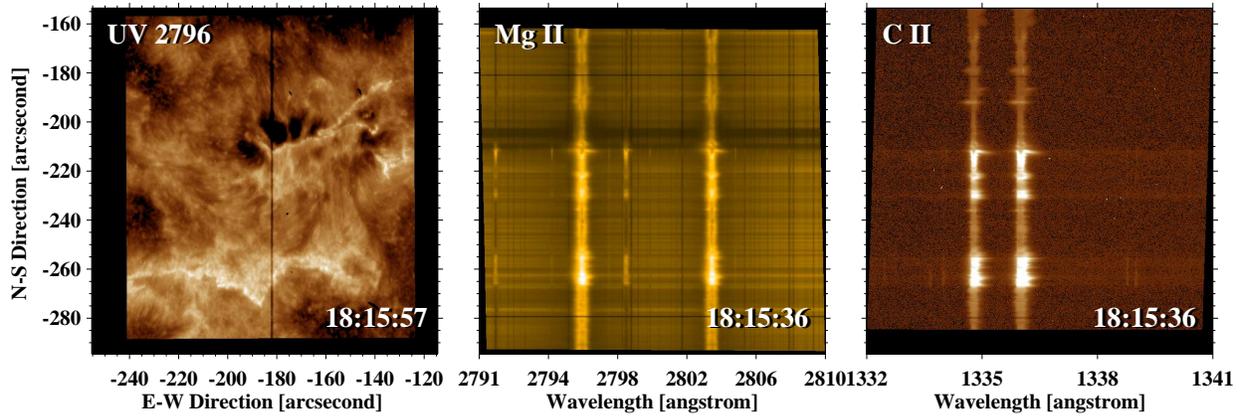}
\caption{IRIS Slit Jaw image (left) and spectra (middle: \ion{Mg}{2}, right: \ion{C}{2}) for the event observed on 2014 August 1 around 18:15 UT.
\label{2ndspe}}
\end{figure}

\begin{figure}[pht]
\centering
\includegraphics[scale=1]{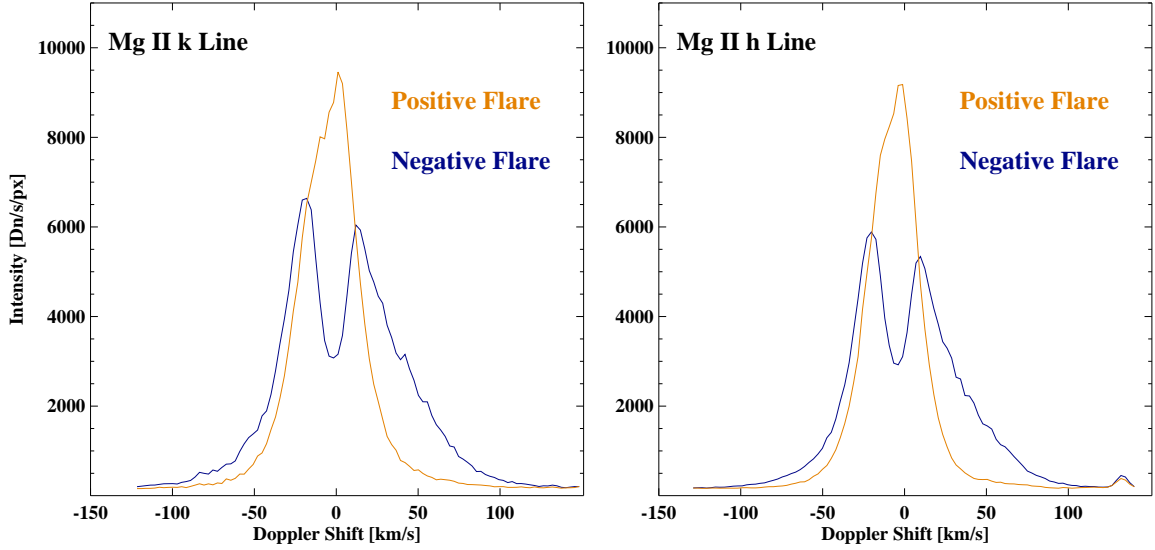}
\vspace{-1.2em}
\caption{\ion{Mg}{2} k (left) line and h line (right) profiles in the positive (emission) flare region with the negative contrast front. The blue profiles are selected from the ribbon front, which shows enhanced absorption in \ion{He}{1} 10830~\AA\ at 18:15:36 UT. The orange profiles are selected from the middle of the northern flare ribbon, where emission is seen in both the UV and \ion{He}{1} 10830~\AA. The slit position is shown in the left panel of Figure~\ref{2ndspe}.
\label{spemg}}
\end{figure}

\begin{figure}[pht]
\centering
\includegraphics[scale=1]{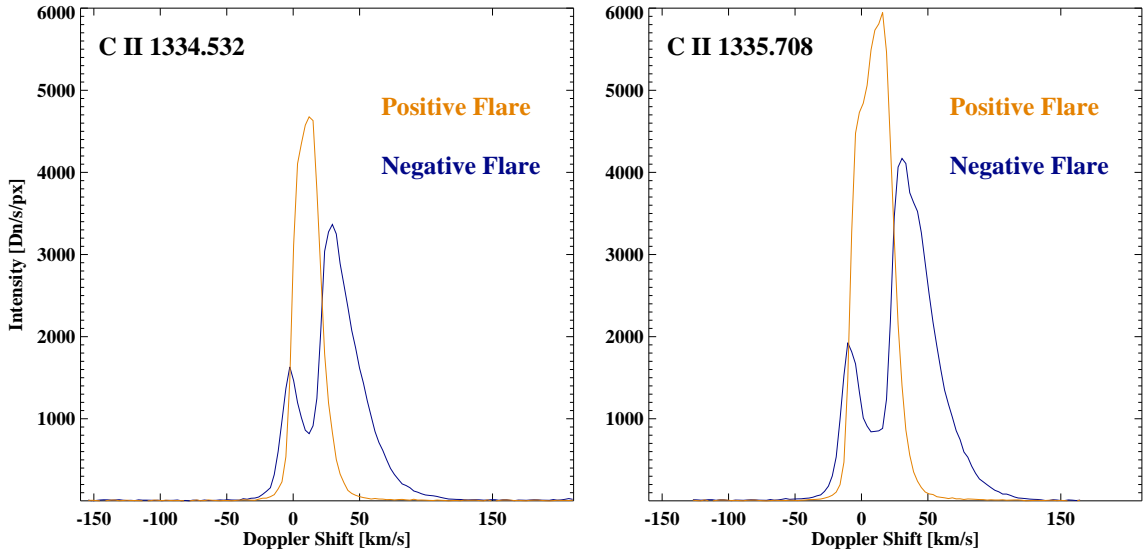}
\vspace{-1.2em}
\caption{\ion{C}{2} line profiles in the positive (emission) flare region with the negative contrast front. The profiles correspond to the same pixel as those in Figure~\ref{spemg}. The slit position is shown in the left panel of Figure~\ref{2ndspe}.
\label{spec}}
\end{figure}

At about 18:15 UT, the IRIS spectral slit was placed at a position that cuts through the flare ribbon perpendicularly, which is perfect to study the different components of the flare ribbon (see Figure~\ref{2ndsji}). In Figure~\ref{2ndspe}, \ion{Mg}{2} and \ion{C}{2} spectra of this slit position are plotted. Recall that the negative flare edge shows negative contrast in \ion{He}{1} 10830~\AA\ and the following part is in emission. Therefore, we anticipate that we will see different spectral characteristics at the front edge of the ribbon compared to the areas behind. Two line groups, the \ion{Mg}{2} lines at  2796.33~\AA\ and 2803.51~\AA\ and the \ion{C}{2} lines at 1334.53~\AA\ and 1335.71~\AA, are studied. The line profiles, of \ion{Mg}{2} and \ion{C}{2}, in representative pixels of the front (corresponding to the dark edge in \ion{He}{1} 10830~\AA) and the middle (emission in all UV lines and \ion{He}{1} 10830~\AA) of the northern flare ribbon, are plotted in Figure~\ref{spemg} and Figure~\ref{spec}. As expected, the front edge (blue) of the northern ribbon shows significant abnormal features compared to the rest of this ribbon body (orange), which is expected because negative contrast is observed in \ion{He}{1} 10830~\AA. Several important characteristics are inferred: 1) In general, the spectra of the ordinary flare ribbon behind the front are brighter than those on the negative front. 2) Strong, mainly red-shifted emission and line broadening (with a FWHM near or greater than 0.8~\AA) are found in the negative front relative to the ribbon behind. 3) Both the \ion{Mg}{2} and \ion{C}{2} show a central reversal pattern in the ribbon front. This is a common feature for quiet Sun spectra \citep{Leenaarts2013} but unusual for flare regions \citep{Kerr2015}.

\begin{figure}[pht]
\centering
\includegraphics[scale=0.8]{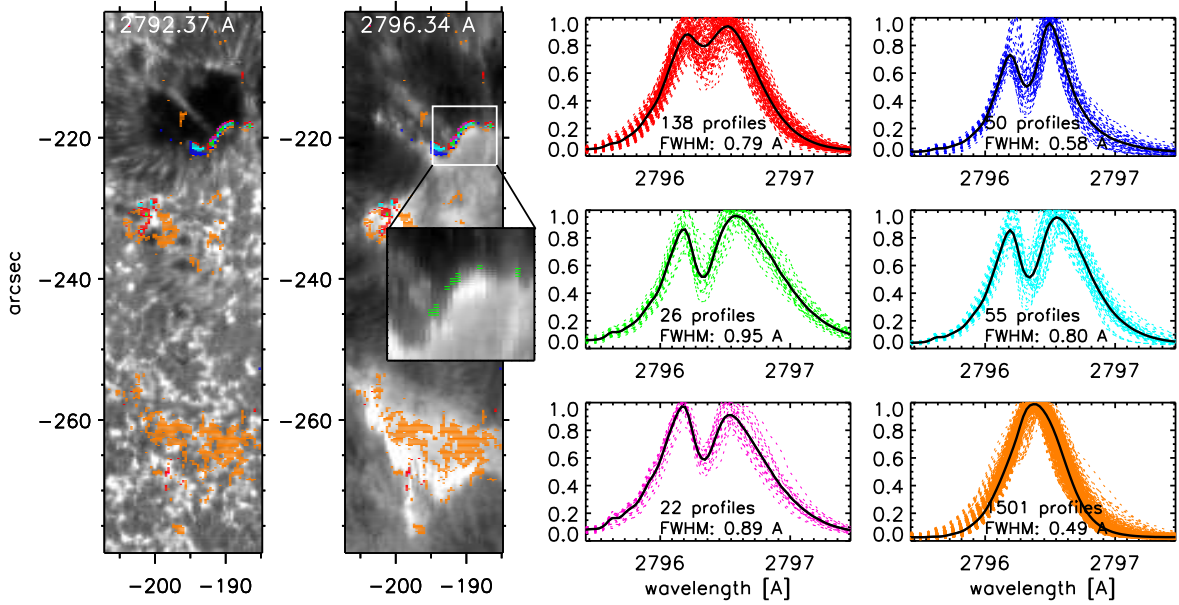}
\vspace{-1.2em}
\caption{\ion{Mg}{2} raster images and line profiles in the northern flare region of the 2014 August 1 event. The raster images (left two panels) were constructed from 17:54:55 to 18:28:55 UT with 64 steps of the spectrograph slit. Using a machine learning method,  \ion{Mg}{2} line profiles are grouped by their shapes. In the right two columns, six representative groups are plotted in different colors. For instance, the profiles in orange color are widely distributed and associated with the southern flare ribbon and some dispersed brightenings. The broadest profiles are in green color, with a FWHM of ~1~\AA. As shown in the magnified inset (second panel from left), these profiles lie in the front edge of the northern ribbon, corresponding to the dark ribbon observed in \ion{He}{1} 10830~\AA.
\label{ml}}
\makeatletter
\renewcommand{\thefigure}{S\@arabic\c@figure}
\makeatother
\end{figure}

Furthermore, we classified the \ion{Mg}{2} profile shapes into groups of self-similar profiles using a machine learning algorithm \citep{MacQueen1967}. A total of 30 groups are classified according to their characteristics, for instance the line broadening and the central reversal. The northern ribbon, with a dark front observed in \ion{He}{1} 10830~\AA, appeared with a special type of \ion{Mg}{2} profile, including the broadest profiles of the whole FOV. A selection of these profiles is shown in Figure~\ref{ml}. The two images on the left show the IRIS raster in two different wavelengths at 2792.37~\AA\ and 2796.34~\AA. Overplotted are the color-coded locations of special \ion{Mg}{2} profiles, whose shapes are shown in the panels on the right. The average profile of each colored group is shown in black and is used to determine the FWHM. The widest profiles of the FOV with a FWHM of ~1~\AA\ are shown in green and from the magnified inset it can be seen that these profiles lie just north of the bright flare ribbon, corresponding to the dark ribbon observed in \ion{He}{1} 10830~\AA. The southern ribbon did not show any especially broad \ion{Mg}{2} profiles and it also did not show any dark ribbon front in \ion{He}{1} 10830~\AA.

\section{Discussion}

The prevalence of the negative flare ribbons observed in \ion{He}{1} 10830~\AA\ provides important and unique constraints in modeling solar flares. The negative contrasts are detected on the advancing edge of the flare ribbon, representing the footpoints of newly reconnected flare loops according to the `standard' flare model  \citep{Cargill1983,Hudson2011}. In other words, the ribbon with a negative contrast corresponds to the latest electron beams penetrating from the solar corona, where the magnetic reconnection occurred. The bright part of the ribbon behind the dark frontier is the signal of the flare heating immediately following the enhanced line absorption. Such a leading front due to electron penetration has been reported in H$\alpha$ observations as narrow bright features with red-shifted spectra, for instance by \citep{Svestka1980} using the multi-slit spectrograph developed by Lockheed Solar Observatory \citep{Martin1974}.

Spectroscopic observations in H$\alpha$ from NST/FISS and UV lines from IRIS reveal distinct features on the ribbon front from that behind the front, namely a strongly enhanced red wing and line broadening, respectively. In newly formed flare ribbons, the red shift is usually the dominant velocity pattern, which is considered an unambiguous signature of mass motion, possibly due to chromospheric evaporation resulting from the electron precipitation. According to previous modeling, two mechanisms are plausible in explaining the enhanced absorption in \ion{He}{1} 10830~\AA. In the collisional ionization-recombination model, the helium atom is excited, through collisional ionization from the ground state by energetic electrons followed by recombinations to triplet states. Our observation can be explained by this model since we are able to correlate the electron-precipitation site with the dark edge of the flare ribbon in 10830~\AA. On the other hand, the photoionization-recombination model assumes that the helium atom is excited by the EUV radiation \citep{Ding2005}. Any EUV radiation with a wavelength shorter than 304~\AA\ can generate excitation of \ion{He}{1} atoms via the photoionization-recombination process. One recent simulation reproduces a much deeper absorption profile in \ion{He}{1} 10830~\AA\ caused by X-ray and EUV radiation \citep{Allred2015}. In such a scenario, the flare ribbon front is heated by the X-ray and EUV radiation and enhanced absorption is seen. As the heating accumulates, for the region behind the front, emission becomes dominant. Therefore, either model can explain the mixture of negative and positive contrasts observed in one flare ribbon. However, as the soft X-ray and EUV radiation is not so strong in the beginning, and rises less impulsively than the non-thermal electron flux, it is hard to explain the narrowness of the dark ribbon. Also, it does not explain the associated strong red-shifts and broadening. Therefore, we think that the first scenario, collisional ionization-recombination, is more promising in explaining the observed narrow negative flare front.

In conclusion, observations of darkening in \ion{He}{1} 10830~\AA\ can be used to provide  strong constraints in modeling the nonthermal effects of solar flares. By contrast, both collisional ionization-recombination and photoionization-recombination mechanisms can contribute to the emission that appears in the following bright ribbon. The decisive physical parameter in determining whether collisional ionization-recombination leads to an absorption or emission is the local plasma temperature. During a strong flare, the heating is more efficient and the lower layers are heated to a higher temperature very quickly. Under these conditions, it is less likely that a negative flare in \ion{He}{1} 10830~\AA\ will be observed. However, weak flares, like the low M-class flares studied, have a more moderate heating pattern. Therefore, we are able to see the dark flare source before the lower atmosphere is heated to a high temperature. Although the exact value of the critical temperature is unknown at this moment, a hint is provided by the D3 observations, in which the threshold is about 25000 K \citep{Zirin1980}. A more detailed study, especially a correlative numerical simulation, is unavoidably hampered in these events due to the lack of RHESSI hard X-ray coverage, which is needed to provide quantitative measurements of the electron energy distribution. Nevertheless, this is a significant and pioneering observation providing a rare but promising opportunity to  understand the non-thermal effects in solar flares. Furthermore, this unique NST observation provides a solid scientific cornerstone for the future observations with the 4-meter Daniel K. Inouye Solar Telescope that will have even higher resolution.

We would like to thank the referee for valuable comments. The data used in this paper were obtained with NST at Big Bear Solar Observatory, which is operated by New Jersey Institute of Technology. Obtaining the excellent data would not have been possible without the help of the BBSO team. BBSO operation is supported by NJIT, US NSF AGS-1250818 and NASA NNX13AG14G, and NST operation is partly supported by the Korea Astronomy and Space Science Institute and Seoul National University, and by strategic priority research program of CAS with Grant No. XDB09000000. IRIS is a NASA Small Explorer mission developed and operated by LMSAL with mission operations executed at NASA Ames Research center and major contributions to downlink communications funded by the Norwegian Space Center (NSC,Norway) through an ESA PRODEX contract. The supportive white light data is taken by SDO/HMI. This work is supported by NSF grants AGS-1153424, AGS-1250374, AGS-1408703, AGS-1348513, and NASA grants NNX-11AQ55G, NNX-13AF76G, NNX-13AG13G and NNX-11AO70G to NJIT. W. Cao was supported by NSF AGS-0847126 and NSFC-11428309. H. Ji was supported by CAS Xiandao-B with grant XDB09000000, NSFC-11333009, NSFC-11173062, NSFC-11221063, and the 973 program with grant 2011CB811402. The work of K.-H. Cho and J. Chae was supported by the National Research Foundation of Korea (NRF - 2012 R1A2A1A 03670387).

\bibliographystyle{apj}

\end{document}